\documentclass[12pt,preprint]{aastex}
\usepackage{natbib}
\newcommand{\el}{$\ell$}
\newcommand{\Pone}{$P_1$}
\newcommand{\Ptwo}{$P_2$}
\newcommand{\Pthree}{$P_3$}
\newcommand{\Ptime}{$P_{time}$}
\newcommand{\Pnode}{$P_{node}$}
\newcommand{\gmodes}{{\it g-}modes}
\newcommand{\gmode}{{\it g-}mode}

\shorttitle{Non-radial Pulsations in Pulsars}
\shortauthors{Clemens and Rosen}
\received{2004 January 29}
\begin{document}

\title{Observations of Non-radial Pulsations in Radio Pulsars}

\author{J. Christopher Clemens\altaffilmark{1} \& R. Rosen}
\affil{Department of Physics and Astronomy, University of North 
Carolina, Chapel Hill, NC 27599-3255}
\email{clemens@physics.unc.edu; rrosen@physics.unc.edu}

\altaffiltext{1}{Alfred P. Sloan Research Fellow}

\begin{abstract} 


We introduce a model for pulsars in which non-radial oscillations of high 
spherical degree (\el) aligned to the magnetic axis of a spinning neutron star reproduce 
the morphological features of pulsar beams.   In our model, rotation of the pulsar carries a 
pattern of pulsation nodes underneath our sightline, reproducing the longitude stationary 
structure seen in average pulse profiles, while the associated time-like oscillations 
reproduce ``drifting subpulses''---features that change their longitude between successive 
pulsar spins.  We will show that the presence of nodal lines can account for observed 
$180^\circ$ phase jumps in drifting subpulses and their otherwise poor phase stability, 
even if the time-like oscillations are strictly periodic. Our model can also account for the 
``mode changes'' and ``nulls'' observed in some pulsars as quasiperiodic changes 
between pulsation modes of different (\el) or radial overtone ($n$), analogous to 
pulsation mode changes observed in oscillating white dwarf stars. We will discuss other 
definitive and testable requirements of our model and show that they are qualitatively 
supported by existing data. While reserving judgment until the completion of quantitative 
tests, we are inspired enough by the existing observational support for our model to 
speculate about the excitation mechanism of the non-radial pulsations, the physics we can 
learn from them, and their relationship to the period evolution of pulsars.


\end{abstract}

\section{INTRODUCTION}
\label{intro}

	Upon the discovery of radio pulsations from pulsars by \citet{hew68}, 
\citet{rud68}~ proposed that the pulses arose from non-radial oscillations of a neutron 
star.  This idea was quickly displaced by a rotational model \citep{gol69}, but 
\citet{dra68}~ again raised the possibility of pulsations when they measured individual 
pulse sequences for two pulsars and found within them narrow subpulses that moved to 
successively earlier times within the main pulse.  Because this drift represented the 
presence of a ``second periodicity'' incommensurate with the spin period, it was natural 
to propose a time-like oscillation of the star.  Subsequent measurements, however, 
revealed complex subpulse patterns that did not conform to a pulsation model in any 
obvious way. Moreover, the persistence of unique subpulse shapes from pulse to pulse, 
along with problems of phase stability we will address in later sections, led Drake to 
conclude that the drifting subpulses were incompatible with the pulsation hypothesis 
\citep[see][]{sta70,hew70}. Ultimately, pulsations were abandoned in favor of purely 
geometric models, although they reappeared from time-to-time in the theoretical 
literature \citep[notably][]{hans80,vanh80,mcd88first,car86,fin90,rei92,str93}. Most 
recently \citet{dun98}~ invoked toroidal modes to account for oscillations of soft gamma 
repeaters, but other than the work of \citet{str92}~ and \citet{scvfirst92}, there has been 
no determined attempt to account for the properties of classical pulsars with models 
involving non-radial pulsations.

Instead, most current models, though not all \citep[cf.][]{lyn88,han01}, 
incorporate a circulating pattern of sub-beams, whose motion about the magnetic pole 
produces the drifting subpulses.  In these models, pulsar emission comes from accelerated 
particles that originate near the pulsar magnetic pole and travel along curved paths in the 
star's magnetic field \citep[see][]{rad69,kom70}.  The radiation is confined to a narrow 
beam by the dipole magnetic field geometry \citep{gj69} and relativistic beaming along 
the direction of particle motion, which is roughly parallel to the magnetic axis, not 
perpendicular as in the models of \citet{gol69}, \citet{smi70} and \citet{zhe71}.  The 
observed brightness of pulsar beams effectively demands that the radiation is coherent, 
but the question of how it is produced is not settled \citep{jes01,les98,mel95}.   

Early studies of pulsar single pulses and average pulse shapes \citep[][and 
others]{tay75first, lyn71}~ led to the addition of more elaborate emission structures 
within the model pulsar beam. These features sweep past our sightline and recreate the 
variety of pulse shapes we observe.  \citet{bac76}~ described a target-shaped emission 
pattern (a central core surrounded by an annulus) that can reproduce a wide variety of 
pulse morphologies depending on whether our sightline crosses the center of the pattern, 
yielding a three component pulse, or crosses only the annulus, resulting in a one or two 
component pulse.  \citet{ost77} added a second annulus and rotating features to 
reproduce pulses with more than three components and drifting subpulses.  In 1975, 
\citet{rud75} supplied a physical basis for the model by suggesting that the emission 
arises from localized discharges or sparks near the polar cap. These are arranged in 
annular patterns, and rotate naturally due to the crossed components of the magnetic and 
electric fields.  

In addition to the fixed and drifting substructure, models must account for 
observations of two kinds of discrete events observed in some pulsars; ``mode changes'', 
which abruptly alter the character of the substructure, and pulse ``nulling'', during which 
the pulse emission drops below detectable levels for one or more spin periods of the 
pulsar \citep{bac70a,bac70b,bar82}.   In the  \citet{rud75} model, mode changes and 
nulling result from a collapse or reorganization of the fixed and moving spark structures, 
after which they must reappear with the same features they had previously.  

Several reviewers have summarized observational and theoretical progress in the 
study of pulsar beams.   The most ambitious is Rankin 
\citep{ran83a,ran83b,ran86,ran90,rad90,ran93,mit02}, who has both reviewed and 
synthesized the observations into an empirical model incorporating polarization and 
spectral behavior.  \citet{man95} gives a somewhat different view of the beam geometry.  
Most recently, \citet{gra03} has published a succinct review that includes both ``normal'' 
and millisecond pulsars.

Against this backdrop, as a student project, we conducted a re-analysis of archival 
data on PSR0943+10 to look for evidence of non-radial pulsations, which, according to 
theory, might have periods ranging from milliseconds to seconds \citep{mcd88}.   Our 
analysis, which will appear in a subsequent paper, convinced us that time-like oscillations 
with a period of $31.8$ msec are a viable alternative to the rotating carousel of emission 
beams proposed by \citet{des01}, but we could find no compelling reason other than 
aesthetics to~{\it prefer} a pulsational model.  In search of a definitive test, we reviewed 
the extensive observational literature on pulsars, and found intriguing evidence for 
non-radial pulsations as a universal mechanism for drifting and stationary subpulses.  
Moreover, we found that the original reason for abandoning pulsational models does not 
apply to non-radial pulsations of high azimuthal degree (\el) in which our sightline 
crosses pulsation nodal lines.  The presence of nodal lines increases the variety and 
subtlety of expected subpulse behavior.  

The purpose of this paper is to introduce a model in which high \el~ pulsations 
aligned to the pulsar magnetic pole take the place of the fixed and moving structures of 
the circulating spark model, but other details of the geometry remain unchanged.  In this 
paper we will explore only the phenomenological consequences of this substitution, and 
compare them qualitatively to published observations.  We will not discuss in any detail 
problems in the physics of pulsed radio emission or polarization mechanisms.  In 
\S\ref{morph}, we will present the basic features of our model, and explore its 
observational properties, some of which are not immediately obvious.  Our main purpose 
is to lay the groundwork for future application of the model to radio measurements of 
individual and average pulse profiles.   In \S\ref{comp} we will examine qualitative 
evidence in favor of our model, reserving quantitative comparisons for subsequent 
papers.  The strongest evidence we will present comes from published measurements that 
show subpulse phase behavior difficult or impossible to explain using the circulating 
spark model, but demanded by high \el~ pulsations. We will also discuss analogies 
between pulsar behavior and that of known pulsating stars, specifically the rapidly 
oscillating peculiar A stars (roAp) and the pulsating white dwarf stars. This will 
demonstrate that there are precedents for the model behavior we propose.  In \S\ref{disc}, 
we will speculate about theoretical aspects of our model, such as the pulsation driving 
mechanism, and we will introduce the notion of ``horizontal mode trapping'', which can 
account for the high \el~ character of the proposed modes and relate them to the observed 
period evolution of pulsar beam widths.  We will end by highlighting the potential for 
neutron star seismology, which can yield direct measurements of interesting physical 
quantities like the buoyancy of neutron star surface oceans.

\section{MORPHOLOGY OF NON-RADIAL PULSATIONS}
\label{morph}
In this section we will describe the basic emission patterns that we expect 
non-radial pulsations to produce.  It is simplest, though not strictly necessary, to confine 
ourselves to pulsations where the material displacements 
follow spherical harmonics with azimuthal order $m = 0$.  Using the notation 
\citet{rob82} applied to white dwarfs, we can express displacements as follows:

\begin{equation}
\label{osc}
\xi = Y_{\ell,0}(\Theta, \Phi) \cos(w_tt+\phi) 
\end{equation}

Where the $Y$ is a spatial distribution of pulsation amplitudes, $\Theta$ and $\Phi$ are 
spherical coordinates aligned to the {\it magnetic} axis of the star, and $\cos(\omega_t t 
+\phi)$ is a time-like variation. 

\begin{figure}
\caption{Two sequences of 100 pulses each from 430 Mhz observations of 
PSR0943+10 (lower panels), and their averages (upper panels). The 
subpulses show organized drift from right to left, along with 
disorganized amplitude behavior, but their averages converge to similar 
envelopes.  The data are from \citet{sul98} and are also published in 
\citet{des01}.} 
\label{0943}
\end{figure}

\subsection {Fixed and Variable Pulse Structure}

Already we see in equation 1 the expression of an important feature of pulsar 
emission.  In a series of individual pulses from a pulsar there may be no two alike, yet the 
average of a sufficient number of pulses builds up a profile that is stable in longitude and 
repeatable.  Figure~\ref{0943} illustrates this behavior with two sequences of pulses 
from pulsar PSR0943+10 and their respective averages. A series of ``driftbands'' that 
represent the positions of subpulses in successive spins of the pulsar, can be seen going 
from right to left.  Note that within each individual pulse there are two, sometimes three, 
subpulses, but the average pulse shape is single-peaked.  The abscissa is actually time 
within the pulse, but as conventional we plot longitude calculated according to the 
formula $\Phi = 360^\circ/P_1$, where \Pone~ is the pulsar spin period.  Following 
standard convention, we will use \Ptwo~to represent the time interval between 
consecutive subpulses (the horizontal spacing between subpulses in figure~\ref{0943}) 
and \Pthree~to represent the time required for subpulses to return to a fiducial longitude 
(the vertical spacing between subpulses in figure~\ref{0943}).

Though they exhibit large pulse-to-pulse amplitude variations, the drifting pulses 
in figure~\ref{0943} are modulated, on average, by a longitude stationary envelope. In 
some pulsars this has more complex structure than in PSR0943+10, as we shall see in a 
moment.  In her review and synthesis of pulsar data, \citet{ran83a} expressed the 
difference between the information carried by average pulsar profiles and sequences of 
individual pulses thus:  ``it seems that profiles and pulse sequences must then each 
manifest some largely independent physical basis in the emission region.''  As seen in 
equation 1, non-radial oscillations offer a natural separation between fixed and variable 
structure in the form of spatial and time-like portions of a normal oscillation mode.

In addition to oscillations obeying equation 1, we must also propose that these 
oscillations are coupled to the radio emission mechanism, and that they can generate 
time-modulated emission according to $\xi$ of equation 1. For consistency with the 
observations, we do not want the pulsations to subtract emission in the negative part of 
their cycle, so in our simulations we have kept only positive values of $\xi$.  An alternative 
approach, analogous to the method of \citet{edw02}, would be to add a longitude-dependent 
bias to $\xi$.  This would change the appearance of the simulated individual and average pulse shapes, 
and if large enough, would mute the nodal structure in the average pulse shapes relative to those 
shown in this paper.
  
We also assume that for a fixed radio frequency band, the emission 
originates at about the same altitude above the magnetic pole.  This means that 
longitudinally distinct regions on the stellar surface will correspond to longitudinally 
separated pulse components, though the separation grows larger with increasing emission 
altitude due to the dipole field geometry \citep{kmc70,tho91}.  This assumption is 
consistent with the measurements of \citet{gil91} and \citet{gil92}, who found similar 
emission heights for the various pulse components, but in conflict with the picture 
described by \citet{ran93} or even \citet{gan01} \citep[see also][]{gup03}. 
Finally, we note the sinusoidal pulses that result from 
equation 1 will not be sufficient to reproduce the non-sinusoidal profiles seen in 
figure~\ref{0943}, or the large variations in pulse size, but they will illustrate the 
essential morphological features pulsations can produce.  This level of abstraction will 
allow later incorporation of simulated emission mechanisms \citep[e.g. shot-noise 
models,][]{ric75,scv92}  without affecting the tests of our model presented in this paper. 

Figure~\ref{model}~ shows oscillations with $\ell = 70$ and $m=0$ mapped onto 
the surface of a neutron star aligned with the magnetic axis. Dark regions indicate 
negative displacements and light regions positive ones.  After a half cycle of the 
pulsations, the dark regions would be light and {\it vice versa}, but the nodal lines 
separating them would remain unchanged, except for rotation of the whole pattern about 
the rotation axis of the star (shown as a line extruding from the top of the sphere in 
figure~\ref{model}).  This model is similar to the ``oblique pulsator'' model developed 
for roAp stars by \citet{kur82}, except that \el~is much higher here, and the pulsar only 
emits from a small region near the magnetic pole.  We have indicated the boundary of the 
emission region in figure~\ref{model}~ with a circle around the magnetic pole.  In 
\S\ref{disc} we will present a justification for why this boundary should coincide with a 
nodal line, and we will propose that the oscillations have different amplitude, perhaps 
even zero amplitude, outside of this boundary, a property we have not tried to reproduce 
in figure~\ref{model}.  Following convention, we will use $\alpha$ to denote the angle 
between the pulsar's spin and magnetic axes, and $\beta$ for the minimum angular 
separation between the magnetic pole and our line of sight, which is sometimes also 
called the ``impact parameter''. 

If the pulsar in figure~\ref{model}~ rotates such that the emission region passes 
under our line of sight, we can observe two different kinds of variations.  Because the 
oscillation amplitude is always zero at nodal lines, but can be non-zero elsewhere, the 
nodal lines sweeping past our line of sight can create pulses with a repetition rate related 
to the rate of nodal line passage.  To estimate this rate, consider the case $\alpha = 90$,  
$\beta = 0$. In one full spin of the star our sightline crosses each nodal line twice, so the 
crossing rate is $P_1/2\ell$, but the period of a full cycle of the variations is twice this 
amount, or $P_1/\ell$, because $\xi$ changes sign at each nodal line. For arbitrary 
$\alpha$, the number of crossings is reduced by $\sin(\alpha)$, so the apparent average 
period of the spatial variations is:

\begin{equation}
\label{epnode}
P_{node}  = {{P_1}\over{\ell sin(\alpha)}}
\end{equation}

\begin{figure}
\caption{An oblique pulsator model for pulsar beams, showing an $\ell = 
70, m = 0$ spherical harmonic aligned to the magnetic axis of a neutron 
star.  The angle between the rotation axis and magnetic pole is $\alpha 
= 50^\circ$ in this illustration.   The circle around the magnetic pole 
in the enlarged view denotes the boundary of the emitting region.  This 
region is crossed by four sightlines with different impact parameters 
($\beta$, see text).  For each sightline, the inset shows the 
corresponding rectified slice of the spherical harmonic, representing 
the average beam profile.  At the boundaries corresponding to nodal 
lines, subpulse phase changes by $180^\circ$, denoted by alternating 
$+$ and $-$ signs in the figure.} 
\label{model}
\end{figure}

For simplicity, we have suppressed the more complicated dependence on $\beta$, which 
can be seen in the inset of figure~\ref{model}. The important feature to recognize is that 
the zeroes caused by the spatial node pattern remain at fixed longitude in subsequent 
spins of the pulsar unless either \el~or the emission geometry changes.  

At the same time as these nodal lines sweep past, the time-like oscillations 
generate pulses with a repetition rate related to the oscillation frequency as follows:

\begin{equation}
\label{eptime}
P_{time}= {{2\pi}\over{\omega_t}}
\end{equation}

The behavior we observe in a pulsar beam depends upon the relationship between 
these two periods.  If $P_{node} > P_{time}$, then we will see subpulses narrower than 
the nodal line structure, and, as long as \Ptime~ is incommensurate with \Pone, these 
subpulses will drift in longitude. Furthermore, as long as the measurement does not span 
a nodal line, the separation between subpulses \Ptwo~ will be approximately equal to 
\Ptime.  \Ptwo~ is not exactly \Ptime~ because the nodal structure that modulates the 
amplitudes of the subpulses also affects their times-of-maxima.  For $P_{node} > 
P_{time}$, this causes longitude dependent subpulse drift such that \Ptwo~ is less than 
\Ptime~ near nodal lines.  In Appendix A we quantify this behavior and show examples of 
the driftband curvature it generates. 

In contrast to the appearance of individual pulses, the average of a sufficiently 
large number of pulses will reveal the fixed nodal line structure.  In the inset of 
figure~\ref{model}, we have shown what this nodal line structure would look like by 
plotting various traverses our sightline might make across the magnetic pole.  For each 
traverse, we have plotted a rectified spherical harmonic to simulate the average of many 
spin periods where emission occurs only when $\xi$ is positive in equation 1. Our figure 
is intentionally similar to that of \citet{bac76}, but whereas the spacing and width of his 
annular features was arbitrary, ours follows the spacing and shape of spherical harmonics.  
We will return to this and other features after considering the case where $P_{node} < 
P_{time}$.

For $P_{node} < P_{time}$, individual pulses show no structure significantly 
narrower than the nodal line spacing, but the modulation of fixed pulse components at 
\Ptime~ can still generate quasi-stationary driftlike variations.  We have described the 
approximate behavior of pulse maxima for $P_{node} < P_{time}$, in Appendix A, and 
have shown an example of synthetic data for this case in the right hand panel of 
figure~\ref{nodes}.  Measurements of \Ptwo~ from a single pulse in this case will be 
strongly affected by \Pnode, making it difficult to estimate \Ptime~ without modeling.  In 
spite of these differences, the average of a large number of pulses will look the same as in 
the case previously discussed, and as simulated in the inset of figure~\ref{model}. 

\begin{figure}
\caption{ Oblique pulsator simulations representing individual and 
average pulse profiles of PSR0943+10 (left) and PSR1237+25 (right).  
The PSR0943+10 simulation uses $\ell=83$, $\alpha = 11.5^\circ$, $\beta 
= 5.4^\circ$, $P_1 = 1.098$ s, and $P_{time} = 31.78$ msec.  There are 
no nodal lines in the pulse window, and $P_{time} < P_{node}$, so 
subpulses appear to drift continuously across the profile. The 
PSR1237+25 simulation uses $\ell=85$, $\alpha = 53^\circ$, $\beta = 
0^\circ$, $P_1 = 1.382$ s, and $P_{time} = 89.90$ msec.  There are four 
nodal lines in the pulse window, and $P_{time} > P_{node}$ so subpulses 
appear as quasi-stationary variations with phase reversals at the nodal 
lines.} 
\label{nodes}
\end{figure}

For both cases, there is a $180^\circ$ shift in subpulse phase between pulse 
components separated by a nodal line.  This has been indicated by alternating $+$ and $-
$ signs in the inset of figure~\ref{model}.  This means that the driftbands caused by 
drifting subpulses, like those in figure~\ref{0943}, will not be continuous across nodal 
lines.  We have simulated this behavior in figure~\ref{nodes}.  The left hand panel shows 
a model representing PSR0943+10, where the sightline traverse resembles the $\beta = 
3.4^\circ$ case (with outer components missing), or the $\beta = 5.7^\circ$ case shown in 
figure~\ref{model}.  No nodal line is crossed, and the drift is continuous across the whole 
profile.  On the right is a model representing the 5-component profile of PSR1237+25, 
whose impact parameter is smaller.  For this model, adjacent pulse components have 
different driftband phase, so there is no continuous pattern extending across the profile. 
The model we have used to represent PSR1237+25 also has $P_{node} < P_{time}$, 
and illustrates the nature of the drifting in that case.

\subsection {\it {Requirements of the Model}}

The model properties described so far are broadly consistent with the observed 
behavior of pulsar radio emission, but to focus the discussion onto specific tests, we will 
state as succinctly as possible three definitive requirements of the pulsation model for 
comparison with observations.

\newcounter{saveenum}
\begin{enumerate}
\item{At the nodal line separating adjacent pulse components, subpulse 
amplitudes should be zero, and their phase should jump by $180^\circ$.}
\setcounter{saveenum}{\value{enumi}}
\end{enumerate}

This assumes that only one pattern of nodal lines is present at a time, an assumption that 
could be violated if several pulsation eigenmodes are excited simultaneously, as occurs in 
the white dwarf stars.  Note that 1 does not require the radio emission be zero at nodal 
lines, but rather that the modulated component of the emission be zero; we have not 
explicitly required that all of the emission come from the pulsations. As a corollary to 
requirement 1, subpulse phase should drift almost linearly between the $180^\circ$ 
jumps, to within the effects of relativistic aberration and delay \citep[see][]{gil91}. If, 
however, the subpulse phase is inferred from the times-of-maxima of individual 
subpulses, these will follow the curvature calculated in Appendix A.  The literature on 
driftband curvature \citep[e.g.][]{wri81,kri80}~ does not account for the possibility of 
$180^\circ$ phase jumps, but we will show in \S\ref{comp} that they have been observed 
in a number of stars, most recently and dramatically by \citet{esl03}. 

\begin{enumerate}
\setcounter{enumi}{\value{saveenum}}
\item{The spacing between fixed pulse components should follow the same 
distribution as a spherical harmonic sampled along a single sightline.}
\setcounter{saveenum}{\value{enumi}}
\end{enumerate}

This requirement has to incorporate the effects of $\beta$, which is the first of 
several complications.  The second complication lies in the radio frequency dependence 
of average profiles, which is far from simple \citep{mit02}.  It is possible to understand 
these profile dependencies in the context of a radius to frequency mapping model, as first 
proposed by \citet{kmc70} and explored by \citet{tho91}.  In this model, lower 
frequency observations measure emission from a higher altitude, where the dipole field 
has diverged more.  Since the emission is apparently tangent to the magnetic field, this 
divergence introduces a frequency dependent ``magnification''.  This magnification 
broadens pulse components at low frequencies compared to their higher frequency 
counterparts, and changes the  $\beta$ of effective sightline, since the particles emitting 
at higher altitude originated closer to the magnetic pole.  However, if the magnification 
follows a dipole scaling, the ratios between component spacings will be preserved.   
Another difficulty arises from the Gaussian shape \citep{kra94} of measured pulse 
components, whose half-widths will differ from the cosine-like nodal regions of our 
model. We can mitigate this by comparing our model to measurements of pulse 
component maxima instead of widths, when possible. Finally, the emission we see 
probably represents an integral over some area on the star, due to the finite radio 
bandwidth and perhaps divergence of the emission itself.  Together these problems make 
definitive tests problematic, but we will show that the average beam geometries explored 
by \citet{ran90,ran93}, \citet{gou94}, and \citet{gil93}, are crudely compatible with the 
requirements of our model. 

\begin{enumerate}
\setcounter{enumi}{\value{saveenum}}
\item{Within the same pulse component, subpulses follow the relationship 
\begin{equation}
\label{beat}
{{1}\over{P_3}} = {{{1}\over{P_{time}}} - {{n}\over{P_1}}},
\end{equation} 
where $nP_{time} \approx P_1$.}
\end{enumerate}

This arises because \Pthree~ is simply a beat between \Pone~ and \Ptime~ in our 
model.  This relationship is the same as that given by \citet{sta70}~ for pulsational 
models, except we have substituted \Ptime~ for \Ptwo.  As we have discussed, when 
\el~is sufficiently large that one or more nodal lines appear in the observed pulses, 
\Ptwo~ is not necessarily a good estimator of \Ptime, thus {\it we cannot rule out the 
existence of stable clock based solely on the measured irregularity of \Ptwo}.  According 
to \citet{sta70}, one of the primary reasons for rejecting pulsational models for drifting 
subpulses was the relative instability of \Ptwo~ compared to \Pthree.  In \S\ref{comp}, 
we will answer this objection to pulsation models, thirty-four years late, by reproducing 
observations of PSR1919+21, the first pulsar discovered.  We will see that a model 
satisfying requirement 3 can simultaneously exhibit variations in \Ptwo~ like those 
measured by \citet{dra68}~ and \citet{bac70c}.
  
 	As a corollary to 3, neither \Ptime~ nor \Pthree~ should be affected by the 
radius-to-frequency mapping that broadens \Pnode~ at lower radio frequencies.  So while 
the components of an average profile grow farther apart when observed at low frequency, 
the time-like pulses will not.  Once again, it is crucial to recognize that \Ptwo~ may not 
be a good estimator of \Ptime, especially when \Ptime~ exceeds \Pnode. In that case 
measurements of \Ptwo~ can be dominated by the nodal line structure instead of \Ptime.  
As we will discuss in \S\ref{comp}, measurements of the frequency dependence of 
\Ptwo~ show negligible frequency dependence for those pulsars where $P_{time} < 
P_{node}$, and an increasing frequency dependence as \Pnode~ approaches \Ptime~ 
\citep{izv93,gil02}, consistent with the requirements of our model.

\section{COMPARISON TO OBSERVATIONS}
\label{comp}

\subsection{\it{Subpulse Phase Jumps}}

	A phase jump of $180^\circ$ is not subtle behavior, so if our model is correct then 
this property of drifting subpulses should have been observed repeatedly. Interestingly, 
the first measurement of phase differences between adjacent pulse components came 
relatively early, in \citet{tay75}, but its significance for pulsation models was not 
recognized or pursued.  \citet{tay75} constructed individual time series for each of the 
five pulse components in PSR1237+25, and cross-correlated them.  With the exception of 
the central component, their analysis showed that components adjacent to each other in 
longitude have opposite subpulse phase (see their figure 11).  This behavior required 
\citet{ost77} to place the emission regions on their inner circulating carousel out of phase 
with those on the outer \citep[see figure 13 of][]{ost77}, and led \citet{han80}~ to 
propose a spiral emission pattern. In addition to PSR1237+25, \citet{tay75} found similar 
anti-correlations for the components of PSR0329+54.  Eleven years later, \citet{pro86}~ 
applied the same analysis with better resolution to PSR1919+21, PSR0809+74, and 
PSR1237+25 (again).  All three of these objects show behavior consistent with 
$180^\circ$ jumps in their subpulse phases in at least one radio band.  

\begin{figure}
\caption{ Subpulse amplitude (upper panel) and phase (lower panel) 
envelopes for ~PSR0320+39, reproduced with permission from 
\citet{esl03}.  The upper panel also shows the average pulse shape 
(dotted line).  The subpulse amplitude envelope shows a minimum near 
zero at the same longitude as a $180^\circ$ shift in the phase 
envelope, consistent with the requirements of the oblique pulsator 
model.  The phase envelope is plotted three times representing analysis 
via three different techniques. A $60^{\circ/\circ}$ slope has been 
removed from the phases.}
\label{0320}
\end{figure}

In addition to these four objects, \citet{esl03}~ recently applied their 
two-dimensional fluctuation spectrum technique to PSR0320+39, and found dramatic 
evidence for phase and amplitude modulation like that expected at a nodal line.   In 
figure~\ref{0320}, we have reproduced figure~3 of their paper, which was largely 
responsible for guiding us to the model we are proposing.  As \citet{edw02}~ point out, 
the two-dimensional Fourier transform as they apply it makes use of all the phase 
information in the data to produce phase and amplitude envelopes with high 
signal-to-noise ratio even for modest quality data. In figure~\ref{0320}, the phase 
envelope shows the $180^\circ$ phase shift we expect at a nodal line, and almost linear 
behavior in between (a $60^{\circ/\circ}$ slope has been removed from the data).  

At the same longitude as the phase shift, the subpulse amplitude is near zero, as 
required at a nodal line.  This figure evokes comparisons to figure 14 of \citet{kur90}, 
which shows a similar phase shift in the rapidly oscillating Ap star HR3831 as rotation 
changes the viewing geometry of the pulsation nodal structure.  We note that our model 
requires symmetry in the pulse components which means that a second phase jump 
should appear in the profile of figure~\ref{0320}.  In their subsequent paper 
\citet{edw03}~ detected such a  jump near the right hand edge of the profile.  

\citet{edw03}~ applied a similar analysis to PSR0809+74 at two frequencies, 
with results that are more challenging for our model.  The phase envelopes do not appear 
to be linear, and there are abrupt phase shifts not equal to $180^\circ$.  Since the 
emission we observe is an integral over some frequency range and perhaps over some 
area on the star, we speculate that abrupt changes can be ``washed out'' by these inherent 
limits to the longitude resolution, especially at low frequency where the pulse 
components change their appearance most rapidly \citep{tho91}.  Whether PSR0320+39 
is the lucky exception or the norm will require more data to tell.  At any rate we do not 
think the problems with PSR0809+74 should overwhelm our model, especially when 
compared to the elaborations these phase changes require in the drifting spark model 
\citep{edw03}, but caution and careful modeling will be required.

\begin{figure}
\caption{ A comparison of longitude-resolved cross-correlation maps for 
PSR1919+21.  The left panel shows the cross-correlation of 1420 MHz 
time series data from each longitude with that at a reference 
longitude, reproduced with permission from \citet{pro86}.  The right 
panel shows the cross-correlation map of simulated data using an 
oblique pulsator model with $\ell=100$, $\alpha =45^\circ$, $\beta = -
2.35^\circ$, $P_1 = 1.337$ s, and $P_{time} = 32.01$ msec.  The phase 
reversals at $\sim-30$ and $\sim-8$ msec correspond to the locations of 
nodal lines in the model. Solid contours correspond to positive 
correlations.}
\label{1919xc}
\end{figure}

	For PSR1919+21, the prototype of pulsars and of drifting subpulses, we have 
reproduced a longitude resolved cross-correlation map from \citet{pro86}~ in the left 
hand panel of figure~\ref{1919xc}. This is a contour plot of the cross-correlation of the 
time series at each longitude with that at a reference longitude. The maxima occur at lags 
where the subpulse peaks align and minima where peaks align with troughs. For 
subpulses that drift uniformly from one side of the profile to the other, these diagrams 
should be crossed by bands of continuous slope proportional to the drift rate.  Instead, we 
see sloping bands punctuated by two phase inversions, indicating that at those longitudes 
the subpulses abruptly change their phase.  

To illustrate that pulsations can reproduce this behavior, we have simulated PSR1919+21 
with a model like that shown in figure~\ref{model}.  The model has only five 
parameters, $\alpha$, $\beta$, \Pone, \el, and \Ptime, the values of which are listed in the 
caption to figure~\ref{1919xc}.  First we generated a synthetic light curve using the 
positive values of $\xi$ in equation 1 sampled along a sightline defined by $\alpha$ and 
$\beta$.  For each time sample in the light curve, we changed the longitude by 
$\Delta\Phi = 360^\circ \Delta t/P_1$, and the pulsation phase by $\omega_t\Delta t$, 
where $\Delta t$ is the time resolution.  We eliminated data outside the observed pulse 
window, and constructed individual time series at each longitude.  Finally, we 
cross-correlated these time series using the formula provided in \citet{pro86}, and 
produced the contour map shown on the right hand side in figure~\ref{1919xc}.  We 
chose model parameters based on published values (except for \el), in some cases 
adjusting them slightly to improve the fit, which was done by eye. The model parameters 
should be regarded as illustrative only; no attempt was made to measure the quality or 
uniqueness of the fit.  

In figure~\ref{1919lc}, we show individual pulse profiles for our model of PSR1919+21, 
to emphasize the variations that occur in \Ptwo~ even though the pulsation 
frequency is constant. We have indicated two different measured values of \Ptwo~ 
similar to those given by \citet{bac70c}, neither of which is close to the input oscillation 
period of 32.01 msec.  We conclude that the published measurements of PSR1919+21 are 
qualitatively consistent with a pulsational model incorporating a stable pulsation period 
and high \el. 

\begin{figure}
\caption{ Synthetic individual pulse profiles for PSR1919+21, generated 
using the same model as in figure~\ref{1919xc}.  The left-hand side 
shows subpulse separations similar to those measured by \citet{bac70c}. 
The right-hand side shows a larger number of pulses, making the 
amplitude modulation by nodal lines more apparent.} 
\label{1919lc}
\end{figure}
 
\begin{figure}
\caption{ A comparison of longitude-resolved cross-correlation maps for 
PSR1237+25. As in figure~\ref{1919xc}~ the left panel shows a map 
reproduced from \citet{pro86}~ for data at 408 MHz.  The right panel 
shows the map for simulated data using an oblique pulsator model with 
$\ell=85$, $\alpha = 53^\circ$, $\beta = 0^\circ$, $P_1 = 1.382$ s, and 
$P_{time} = 89.90$ msec.}
\label{1237xc}
\end{figure}

	We performed a second simulation for PSR1237+25, with the results pictured in 
figure~\ref{1237xc}.  Once again the qualitative similarity is encouraging.  In this star, 
more of the profile seems to be ``missing'' than in the previous one.  Our model offers no 
ready explanation for these zones where the pulsed emission disappears, but it seems that 
the  mechanism for generating pulsed emission fails. This failure could also explain the 
asymmetry in the profile of PSR0320+39 that we discussed in conjunction with 
figure~\ref{0320}.  To compound the problem, \citet{rad90} report that the pulsed 
emission seldom appears in the core component, and the polarization there is different, 
neither of which has any obvious basis in our model.  On the contrary, the zone at the 
pole of a spherical harmonic has the smallest area and therefore the largest pulsational 
displacements (to make the surface integral equal to those of other zones), so unless some 
mechanism saturates we expect larger amplitudes from the core.

We can put these issues aside pending deeper investigation of the emission 
mechanism, but observations like those described in \citet{han87}~ offer a more direct 
challenge to our model.  They analyzed PSR1918+19 using the same cross correlation 
technique as shown in figures 5 and 7, and found much different behavior. There is 
evidence in that pulsar for different drift rates in each of the 3 components of the average 
profile.  We see no obvious way to reproduce the diagrams of PSR1918+19 with a single 
pulsation frequency, but we note that the time series were very short (just 27 pulses in 
one case) and the inclination $\alpha$ is unusually small.   More measurements and 
detailed modeling may lead to a solution.   

\subsection {\it {Pulse Components and Their Separations}}

        Now we consider the spacing between pulse components in the average profiles of 
pulsars. If the components are related to the zones of a spherical harmonic, as we 
propose, then they cannot have arbitrary widths and separations. For example, in our 
pulsation model, the angular separation between the pulsation pole and the first nodal line 
is $\sim0.44$ times the separation between the pole and second nodal line, independent 
of \el~as long as \el~is high.  Similarly, first and second antinodes have angular 
separations from the pulsation pole in the  ratio $\sim~0.55$.  We expect the modulated 
pulse components in pulsars with $\beta = 0$ to follow similar relationships.  
PSR1237+25 is an good example, since it has five components and  \citet{lyn88} and 
\citet{ran93b} give it  $\beta\approx 0$ (as measured from rotation of the linear 
polarization angle).   As an illustration, we have plotted in figure~\ref{1237pul}~ the 
average profile of PSR1237+25 along with a $\beta = 0$ sightline through rectified 
spherical harmonic of $\ell = 85$.  Note that the actual \el~ at the surface of the pulsar 
will be higher, depending on how much magnification the dipole field geometry has 
imposed.

      It would be much better, though quite difficult, to compare the spherical harmonics to 
a statistical sample of pulsars with known $\beta$.  Fortunately, \citet{ran90,ran93}, 
\citet{gou94}, \citet{gil93}, \citet{kra94}, and \citet{mit99}~ have studied the ensemble 
properties of pulse shapes and uncovered consistent ratios between the core and annular 
emission components after adjusting for or eliminating the effects of $\alpha$ and 
$\beta$. These ratios are statistical averages suitable for comparison to spherical 
harmonics, but first we will discuss their dependence on \Pone. 

\citet{ran90} found that the angular size of the central emission component or 
``core'' follows a $P_1^{-1/2}$ dependence, when allowance is made for variations in 
$\alpha$.  This dependence is the same as that \citet{gj69} calculated for the size of the 
polar cap delineated by open magnetic field lines, i.e. those lines that do not close within 
the velocity-of-light cylinder at $cP_1/2\pi$.  In the Goldreich and Julian model, this cap 
is the region from which charged particles stream off the spinning pulsar, providing both 
a mechanism for shedding angular momentum, and the possibility for radio emission 
from above the magnetic polar cap.  The $P_1^{-1/2}$ dependence of emission cores 
measured by Rankin suggests that they are related to this Goldreich-Julian polar cap, 
although we should note that \citet{lyn88}~ measured a different \Pone~ dependence 
from Rankin. 

 In addition to the functional dependence, \citet{ran90} noticed that at any fixed 
\Pone, the angular half width of pulsar emission cores was the same as the calculated 
apparent angular size of the Goldreich-Julian cap, e.g. about $2.5^\circ$ for \Pone = 1 s.    
This led her to suggest that the core emission fills the cap near the pulsar surface.  While 
appealing in its simplicity, this explanation requires that the annular emission patterns 
originate at different heights along the last open field lines.  \citet{gil91} has criticized 
this suggestion on observational grounds, but provided no alternative physical reason for 
the emission core to follow the scaling of the magnetic cap.  As we will discuss in 
\S\ref{disc}, our model suggests an explanation if the boundary separating open and 
closed field lines can act as a ``mode trapping'' boundary, which is always coincident 
with a pulsation node.  The size of the core region would then have quantized values that 
scale with \Pone~ according to the to size of the polar emission cap.  

\begin{figure}
\caption{ A comparison between a $\beta = 0$ slice through a spherical 
harmonic of $\ell = 85$ and the average pulse profile of 1237+25 
measured at 320 MHz \citep[see][]{ran86}.}
\label{1237pul}
\end{figure}

Based upon her conclusion that the emission core size depends only on \Pone, and 
upon her measurements of the annular emission regions which show the same \Pone~ 
dependence, \citet{ran93} was able to infer that the ratio of the angular sizes of the inner 
and outer emission cones is 1.32, as measured at the outer half-power points.   In similar 
fashion, \citet{gou94} measured half-widths for the core and annular zones and found 
that the core components are 1.4 times as wide.  

If we compare these measured component ratios to analogous ratios for the 
components of a spherical harmonic observed along a $\beta = 0$ sightline, we find that 
the ratio of the half-widths of the core and annular components in the model is  
$\sim1.49$ independent of \el, or about 6\% higher than the Gould measurement of 1.4.  
The ratio analogous to Rankin's  inner and outer cone widths is not as good a match at 
$\sim1.72$, 23\% larger than Rankin's measurement.  However, \citet{gil93} have found 
evidence for a different angular size of the inner components in five component pulsars, 
and the ratio of this innermost component  with the outermost one is 1.62, again about 
6\% below the 1.72 ratio measured in our model.  In \S\ref{disc} we will discuss how our 
model might lead to more than one quantized value for the inner and outer annuli if 
different numbers of nodal lines are trapped within the polar cap.  As we will discuss, 
there is evidence for this kind of quantization in the core component as well, but it has 
been interpreted as a preferred inclination angle $\alpha$ \citep{ran90}.  For now, even if 
the results are somewhat ambiguous, it is gratifying to be able to make any testable 
predictions at all about the shape of average pulsar profiles.   

\subsection{\it{ The Radio Frequency Dependence of Drifting Subpulses}}

	In the process of addressing model requirements 1 and 2 from \S\ref{morph}, we 
have incidentally shown that the formula in requirement 3 is not necessarily contradicted 
by changes in \Ptwo.  Strict confirmation of requirement 3 will be difficult, because all 
attempts to measure \Ptime~ are badly aliased by the narrow, periodic pulse window, 
although PSR0943+10 looks promising in the studies of \citet{des01}.  The best strategy 
may be to concentrate on the wider profiles measured when $\alpha$ is small, since the 
aliasing will not be as bad.  Fourier methods, especially the two dimensional transform of 
\citet{edw02}, will be indispensable to this effort, while direct measurements of \Ptwo~ 
in single or multiple pulses are misleading, as we have seen in our model of 
PSR1919+21.  

	The radio frequency dependence of \Ptwo~ is another area where the behavior of 
subpulse maxima can be misleading.  Our model requires that \Ptime~ and \Pthree~ be 
invariant with radio frequency, but we have already seen that \Ptwo~ will vary near nodal 
lines, even when \Ptime~ does not (see Appendix A). Consequently, since observing at a 
different frequency changes the apparent longitude of the nodal lines, \Ptwo~ can vary 
with radio frequency even though \Ptime~ remains invariant.  Qualitatively, we expect 
the radio frequency dependence of \Ptwo~ to approach zero when \Ptime~ is much 
smaller than \Pnode~ and no nodal lines are near.  For \Ptime~ approaching \Pnode, the 
change in \Pnode~ with radio frequency will modulate \Ptime~ as well and introduce a 
frequency dependence into \Ptwo.  In the limit of very large \Ptime, the radio frequency 
dependence approaches that of the average components.  

\citet{izv93}~ have studied the frequency dependence of subpulses in four 
pulsars.  They found that \Pthree~ does not change with radio frequency, consistent with 
the requirement of our model.   For \Ptwo, the frequency dependence is in all cases less 
than that of the average profile, also consistent with our model.   For PSR0031-07 and 
PSR1133+16, both of which have continuous driftbands across the profile, the 
frequency dependence of \Ptwo~ is very weak, $\sim \nu^{-0.05}$ and $\sim\nu^{-
0.06}$ respectively, while the frequency dependence of the average profile is about 
$\sim\nu^{-0.3}$.  The other two pulsars show larger frequency dependencies for \Ptwo, 
but one of these is PSR0320+39, which we now know has multiple components separated 
by phase jumps, as shown in figure~\ref{0320}.  This means that the subpulse separation 
is similar to the separation between stationary profile components, and we expect a larger 
frequency dependence in this case.  It will be interesting in future quantitative studies to 
attempt to reproduce the exact radio frequency dependence of the subpulses for 
individual pulsars by tuning \Ptime.  In principle, each independent radio frequency 
measurement offers a separate constraint on \Ptime~ that might be useful in verifying its 
stability with respect to \Pthree.

\subsection {\it{Mode Changes and Nulls}}

A number of pulsars exhibit abrupt changes in their drifting subpulse behavior 
\citep{ran86}~ or their average pulse profiles \citep{bar82}, or both.  In some examples, 
these changes are cyclical or quasi-cyclical, such that the pulsar successively visits each 
of two or three modes (e.g. PSR0031-07).   A much larger number of pulsars 
\citep{ran86}, including many of those with mode changes, undergo ``nulls'', in which 
the radio emission falls below detectable levels.  The interesting property of these 
variations in the context of our pulsation model is the ``memory'' the pulsar must retain 
in order to return to the same states repeatedly.  Inasmuch as pulsations represent 
eigenmodes of the neutron star, their eigenperiods reflect a durable physical structure that 
will vary only secularly as the star cools and slows down its spin.  This means that a 
mode excited to observable amplitude and then damped can return again with essentially 
the same period.  

Among the known pulsators, white dwarfs provide some exceptional examples of 
this behavior.  The hydrogen atmosphere variable (DAV) white dwarf G29-38 sometimes 
oscillates with very large amplitudes in a dominant mode with $\sim610$ s period, then 
changes abruptly to a large dominant mode at $\sim809$ s, and then to very low 
amplitude pulsations with no dominant mode \citep{kle98}.  This mode changing 
behavior suggests the exchange of energy between eigenmodes with different amplitudes 
(for the same energy content), though this has not been established with certainty 
\citep{dzi82,wu01}.  We note the similarity of this behavior to pulsar mode changes and 
nulling.   Pulsar mode changing involves changes in subpulse drift rates and in the mean 
profiles, both of which we might expect for changes between modes of different degree 
\el.   Likewise, changes in the subpulse drift rate only might correspond to changes in the 
radial eigennumber $n$.  Furthermore, the exponential recovery of drift rates after nulls 
in PSR0809+74 \citep{lyn83}, suggests relaxation into a normal oscillation mode after a 
mode interaction.  In the same star \citet{vanl02} found that the pulsar is often, and maybe always,
in a different drift mode immediately following a null, which shows that they two phenomena
are physically related.  While it is possible to interpret this in the drifting spark model
\citep{vanl03}, pulsation mode switching may offer a more natural explanation.

Finally, evidence that nulling and mode changing are global phenomena comes 
from the pulsars with interpulses PSR1822-09 and PSR1055-52. \citet{fow81} have 
observed that the interpulse emission in PSR1822-09 changes its intensity when the main 
pulse changes between burst and quiescent modes. Likewise, \citep{big90} found 
intensity correlations between the interpulse and main pulse in PSR1055-52, and 
suggested non-radial oscillation as a possible mechanism for communication between the 
poles. 

\section {DISCUSSION}
\label{disc}

	Our main purpose in this paper has been to set forth the requirements of a model 
for pulsar beams in which non-radial oscillations of high \el~ replace the {\it primum 
mobile} of drifting sparks in the \citet{rud75} model.  This groundwork will clarify 
future applications of our model to individual pulsars, which we will begin in a 
forthcoming study of PSR0943+10. Although quantitative investigations are required for 
definitive tests, we have presented observational evidence that our pulsation model 
should be an active contender for the attention of observers and theorists alike.  We will 
continue with some theoretical speculation about the nature of the pulsations. 

\subsection{\it{The Nature of the Pulsations}}

	When we invoke non-radial oscillations in our model, we mean any oscillations in 
which time-like variations are accompanied by spatial nodal lines that rotation can carry 
past our sightline. Immediately, this suggests various deformations that might appear in 
the core, crust, or ocean of a neutron star \citep{mcd88}, but we should not rule out other 
possibilities such as oscillations in the magnetosphere above the magnetic pole 
\citep[e.g.][]{ryl78, sta70, sch02}.  The main problem we will encounter in identifying 
the pulsations is that most of the oscillations we can propose have frequencies too high to 
account for pulses with the repetition rate of ~30 msec typical for subpulse periods. To be 
fully general we also should not rule out high frequency oscillations that are ``switched'' 
at low frequency, but we will find little guidance in pulsar literature for models of this 
sort.   

	The two quantities that will assist us in identifying the kind of oscillations our 
model should include are \el~and \Ptime.  In our illustrations, we have used 
$\ell\approx70-100$ to match the width of features in observed profiles, but these profiles 
apparently do not originate at the surface of the neutron star. Rather, they are magnified 
versions from radiation emitted at tangents to the diverging dipole field. We can estimate 
the magnification factor, which we will call $f_\nu$, by comparing the observed profile 
widths to the expected size of the Goldreich-Julian emission cap at fixed period \Pone.  
At 1 second, for a pulsar with radius 10 km,  the latter is $1.7^\circ$, while 
\citet{ran93}~ measured $11.5^\circ$ for the width of the outer annular beam.  Together, 
these yield a magnification of $f_\nu\sim7$ at 1 GHz.  Thus an apparent \el~ of 85 
represents a true \el~ at the surface of $\sim600$.  The dispersion relationship for 
\gmodes~ (and torsional modes) \citet{mcd88}~ requires that $P_\ell$ scale as $1/\ell$, 
so if $\ell=600$ has a period of 30 ms, then we expect the $\ell = 2$ mode with same 
radial overtone to have period $\sim9$~s.  This limits considerably the kinds of 
oscillations we might consider.  

\citet{mcd88} gives periods near 9 s for low \el, low radial overtone ($n\approx 
1$), core \gmodes.  However, these modes require very large energies to excite, and are 
trapped in the core by the solid crust, yielding small amplitudes at the neutron star 
surface. An alternative from \citet{mcd88} are the \gmodes~ that propagate in the  $\sim 
1$m thick electron-degenerate Coulomb liquid ocean overlying the solid crust in 
equilibrium neutron star models \citep{ric82}. The low $n$ ocean \gmodes~ have periods 
near 0.3 s, rather than the 9~s we require, but the dispersion relation for \gmodes~ gives 
longer periods for higher overtone, so the eigenperiod should increase to 9 s for $n 
\approx 30$.  Therefore,  ocean \gmodes~ of $\ell\approx600$, and $n\approx30$ match 
the \el~ and \Ptime~ our model requires.  According to \citet{mcd88}, these modes have 
lower energies than the core modes, and large surface amplitudes.  As in white dwarf 
stars, the material displacements in these oscillations are primarily horizontal because of 
the high surface gravity.  

The ocean \gmodes~ also have an associated temperature variation that offers a 
way to modulate the flow of charged particles from the pulsar surface at the pulsation 
eigenperiod.  If we accept the results of \citet{jes01}, the electrons accelerated along 
open field lines from a pulsar magnetic pole can be provided by thermal and field 
emission from the neutron star surface, without the formation of a vacuum gap where 
sparks  originate in the \citet{rud75} model.    This result practically requires that 
subpulses be related to a thermal variation at the neutron star surface, as \gmodes~ 
provide.  Thus non-radial oscillations appear to satisfy one of the basic requirements of a 
pulsar emission mechanism.

The other requirement for emission is that the liberated electrons be ``bunched''.  
Their acceleration away from the surface is naturally provided by the potential difference 
between the neutron star pole and the circum-pulsar medium \citep{gj69}, but conversion 
of the particle stream to coherent radiation requires bunching of charges, whether the 
conversion is via an antenna or a maser mechanism \citep{mel95}.  The radial structure 
of $n=30$ ocean \gmodes~ consists of $\sim3$~cm zones of alternating pulsation phase.   
At any instant, this will correspond to a periodic variation of the temperature and pressure 
with depth.  In electron degenerate plasma, the temperature is a property of the ions while 
the pressure is a temperature insensitive property of the electrons.  The displacements of 
an ion fluid element generate a buoyant restoring force only through their electrical 
coupling to the electrons, so the vertical pressure variation is associated with a vertically 
varying component of the electric field.  If this can modulate the flow of electrons, as in a 
klystron, then the pulsations might also provide the mechanism for bunching the 
electrons emitted from the surface.  The observational clues necessary to clarify this 
question may lie in subpulse polarization measurements, which are beyond the scope of 
this paper. 

A problem with this picture comes from the work of  \citet{car86}, who added a 
strong magnetic field to the pulsation calculations.  Because of the high electrical conductivity
in the neutron star ocean, \citet{car86} treated the magnetic field as ``frozen-in'', and 
recalculated the pulsation frequencies for $B=10^{12}$~G in the MHD limit.  He found that
the ocean \gmodes~ become ``magneto-gravity'' modes with very short 
periods ($< 1$ms) and a different dispersion relation.  The solution to this 
conundrum is provided by the conductivity calculations of \citet{pot99}.  Potekhin finds
electrical conductivities in the direction transverse to a $B=10^{12}$~G field to be four to
five orders of magnitude lower than the $10^{19}$~s$^{-1}$~ \citet{car86} assumed.  
Thus the ohmic diffusion timescale
for displacements of 10~cm or smaller is $\stackrel{<}{\sim}1$~msec, shorter than subpulse periods. 
Not only 
does this mean the calculations of \citet{car86} do not apply, it also means the magnetic field
can simultaneously provide the driving mechanism, the amplitude limiting mechanism, and the 
mechanism for enforcing high radial overtones.

\subsection {\it{Mode Driving and Trapping}}

	The high \el~ in our model suggests the concentration of pulsation driving energy 
into a small surface patch, otherwise it would average away in the sum over multiple 
surface zones with alternating phase. This concentration suggests consideration of the 
emission pole itself as the site of driving.  The torque exerted by braking from particle 
emission will be concentrated at the open field lines, and communicated to the rest of the 
star by magnetic and mechanical dissipation.   If the magnetic field is coupled (even 
weakly) to the surface and it displaces material laterally, the possibility for feedback and 
mode driving exists.  For example, suppose that torque on the open field lines results in a 
displacement of material on the polar cap.  The heating that results can increase particle 
emission, which increases the torque.  Depending on the time delays, this feedback could 
drive oscillatory motion.  Other possibilities involving direct shaking of field lines 
\citep{bor76} by displaced material are also possible. 

	Whatever the driving mechanism, if there are \gmode~ pulsations in the neutron 
star ocean, it is reasonable to propose that their propagation behavior changes at the 
boundary between the emitting cap and the rest of the star, not only because the magnetic 
field changes from an open to closed configuration, but because the surface boundary 
condition changes. Thus the edge of the emitting cap constitutes a circular boundary that 
might ``trap'' pulsation modes in the horizontal direction, analogous to the trapping by 
composition transition zones in white dwarf stars \citep{win81}.  This ``horizontal mode 
trapping'' could provide a connection between the size of the emitting cap and the size of 
pulsation nodes by forcing surface nodal lines to lie at the circular boundary.   If this 
connection is maintained as \Pone~ increases, it can explain why the core emission zone, 
as measured by \citet{ran90} and others, follows a $P_1^{-1/2}$ relationship. In this 
section, we will explore pulsation period evolution in the context of a horizontal trapping  
model.

\subsection{\it{The Period Evolution of the Pulsations}}

The pulsars measured to date seldom, if ever, have more than five components in 
their average pulse \citep[cf.][although their methodology is compromised by pulsations]{gan01}
  In our model, five components could result from horizontal mode 
trapping at the third nodal line from the pulsation pole, and this is the geometry we 
depicted in figure~\ref{model}.  We have already seen that parts of the profile can be 
``missing'', so the presence of five components in the model does not necessarily mean 
that we see all five.  If we now force \el~ to have a value that places the third node at the 
emission boundary, we can write a relationship for the period evolution of \el.  From 
\cite{gj69}, the width of the emission cap for a 10 km star follows:

\begin{equation}
W_{cap}=1.7^\circ P_1^{-1/2},
\end{equation}
and the width of third nodal line scales as $1/\ell$, reaching  $1.7^\circ$ for $\ell \approx 
600$.  So we may write:

\begin{equation}
W_3 =1.7^\circ{{600}\over{\ell}}.
\end{equation}

Enforcing horizontal mode trapping requires $W_{cap} = W_3$ so,

\begin{equation}
\ell = 600P_1^{1/2},
\end{equation}
as long as the trapping stays at the third nodal line.  This equation is a mathematical 
statement of the obvious requirement that as the emission cap shrinks during spin-down, 
the \el~ of trapped pulsations must go up.  Some of the other consequences for the 
pulsations are not as obvious.  For instance, from the dispersion relation for \gmodes~, 
\Ptime~ changes as $\sim1/$\el~ if the radial overtone doesn't change, so we expect 
\Ptime~ to decrease as \Pone~ increases.  However, the important ratio 
$P_{time}/P_{node}$ becomes:

\begin{equation}
{{P_{time}}\over{P_{node}}}\propto{{\sin(\alpha)}\over{P_1}}
\end{equation}
So as \Pone~ increases (or the spin and magnetic axes align), \Ptime~ shrinks with 
respect to \Pnode~.  This favors the appearance of narrow drifting subpulses in the long 
period pulsars, as observed. The impact parameter $\beta$, which we have ignored, may 
also play a role, since the decreasing emission cap size makes it less likely that our 
sightline intersects the central component (see figure~\ref{model}), and \Pnode~ 
increases for such sightlines.

Finally we consider the possibility that the horizontal trapping may apply to the 
second nodal line instead of the third, requiring a lower value of \el~ for the same \Pone.  
In this case, the central component will have a larger angular size at the same \Pone~ than 
for third node trapping.  This would be observed as a bimodal distribution of core 
components, and there is evidence in \citet{ran90} for just such an effect (see her figure 
1).  Because \citet{ran90} assumed that the distribution of core sizes is solely an effect of 
$\alpha$, the bimodality appears as an excess of pulsars at $\alpha = 35^\circ$.    While 
our explanation would eliminate this puzzling excess, it raises two problems of its own.  
First, we would expect some pulsars with interpulses in the distribution with larger core 
sizes, and \citet{ran90} finds none.  Second, figure 1 of \citet{ran90} shows the beam 
sizes for pulsars with single components, not five, meaning all the pulsars in this sample 
have pulse components within their emission caps that do not appear in the mean profiles.
 
\section{CONCLUSIONS}
\label{conc}

Whether or not the foregoing discussion has revealed anything about pulsars, it 
has certainly demonstrated \citet{cle83} maxim that we can get  ``wholesale returns of 
conjecture out of such a trifling investment of fact.''  Nevertheless, the fact remains:  
pulsar beams show subpulse phase reversals at the longitude-stationary boundaries 
separating individual pulse components. We have shown that these changes are 
comprehensible in the context of an oblique pulsator model incorporating non-radial 
pulsations of high degree \el.  The important features of our model are: 
non-radial oscillations aligned to and symmetric about the pulsar {\it magnetic} axis;
surface displacements that follow a spherical harmonic distribution;
radio emission that follows the displacements but is never negative;
pulsation modes of sufficiently high \el~ that nodal lines often appear in the pulse window;
and pulsation frequencies that remain coherent over many pulsar spin periods.
Variations on this basic model might include multiple pulsation modes with non-zero 
azimuthal orders, pulsations that are distorted, in reality or in appearance, by non-dipole 
fields, and modes that interact either through mode coupling or a non-linear emission mechanism. 
 
Our model qualitatively reproduces the mean shapes of 
pulsar beams and the radio frequency dependent behavior of subpulses with a minimum of free parameters.
In the most basic form of the model these are $\alpha$, $\beta$, \Ptime, \el, and \Pone.   Our model also 
dictates specific requirements that can be tested quantitatively using new or archival data.  
We have embarked on a program to conduct such tests and we encourage others to do 
likewise. If the model survives these tests, then we will have the opportunity to measure 
fundamental properties of matter in a domain not accessible to laboratory experiments.  
The first challenge will be to determine the site of the pulsations, and then to connect 
measured eigenfrequencies with the eigenmodes of a structural model.  Given the number 
of modes in the pulsation spectrum at large \el, this may be a daunting task, but even 
rough identification will provide limits on the thermal, electrical, and mechanical 
properties of constituents of a neutron star, the densest objects accessible to direct 
observational scrutiny.

\acknowledgments

We are grateful to Chuck Evans and Don Winget for helpful 
conversations, and to Joanna Rankin for providing the archival data on 
PSR0943+10.  One of us is grateful to Gus and Jack for providing a 
quiet environment for the completion of this work. This work was 
supported by a CAREER grant from the National Science Foundation (AST 
000-94289) and by a fellowship from the Alfred P. Sloan Foundation.

\appendix

\section{MODULATION INDUCED DRIFTBAND CURVATURE}

     The purpose of this appendix is to calculate, in approximate fashion, the apparent 
change in subpulse drift between nodal lines caused by the amplitude modulation of 
strictly periodic time-like pulses.  We will approximate the longitude-dependent 
modulation between two nodes as a cosine function, which is very similar to the envelope 
between two nodes of a spherical harmonic for sightlines with $\beta = 0$.  Thus we can 
write a more manageable version of equation 1:
\begin{equation}
\label{A1}
\xi = \cos(\omega_\Phi t)\cos(\omega_t t+\phi),
\end{equation}
where $\omega_\Phi = {{2\pi}\over{P_{node}}}$.

To find the times of maxima (and minima), we take the derivative and set it to zero, 
yielding:

\begin{equation}
\label{A2}
 {-{\omega_\Phi}\over{\omega_t + \phi}} = {{\tan(\omega_t t + 
\phi)}\over{\tan(\omega_\Phi t)}}.
\end{equation}
In figure~\ref{drift}~ we have plotted the locus of the times-of-maxima for $\omega_\Phi 
t $ between $-90$ and 90 degrees, simulating the range between nodal lines, for a variety 
of cases.  The lines in these plots are analogous to subpulse driftbands, because they 
show how the maxima (and minima) of time-like pulses vary with longitude.  For the 
case where $P_{time} \approx P_{node}$, we expect no driftband curvature.  When 
$P_{time} < P_{node}$, we expect slower drift near nodal lines (right hand panels), and 
vice versa.  

We note that for $\beta \neq 0$ sightlines that graze along a nodal line in the center of the 
pulse window, the driftband curvature can be the reverse of the cases plotted here, i.e. 
narrow subpulses will drift more slowly near the center of the profile.  This can explain 
the driftband curvature measured for PSR0031-07 by \citet{kri80}, and discussed by 
\citet{wri81}.  The more important conclusion is that when nodal lines are present, 
\Ptwo~ can vary with longitude, even though the underlying clock is absolutely stable.

\begin{figure}
\caption{ Simulated driftband curvature calculated from 
equation~\ref{A2} for cases where $P_{time} = P_{node}$ (top), 
$P_{time} > P_{node}$ (left) and $P_{time} < P_{node}$  (right).}
\label{drift}
\end{figure}

\end{document}